\documentclass{article}

\usepackage{cite,amsmath,amsfonts,amsthm,fullpage}
\usepackage{amsfonts}
\usepackage{amsmath}
\usepackage{color}
\usepackage{graphicx}
\usepackage{amssymb}
\usepackage{amsbsy}
\usepackage{epsfig}
\usepackage{verbatim}
\makeatletter \@addtoreset{figure}{section}
\def\thefigure{\thesection.\@arabic\c@figure}
\def\fps@figure{h, t}
\@addtoreset{table}{bsection}
\def\thetable{\thesection.\@arabic\c@table}
\def\fps@table{h, t}
\@addtoreset{equation}{section}

\newtheorem{corollary}{Corollary}[section]

\newtheorem{examps}{Examples}[section]

\def\bx{\begin{example}}
\def\ex{\end{example}}
\def\bxs{\begin{examps}. \rm\begin{enumerate}}
\def\exs{\end{enumerate}\end{examps}}
\def\bd{\begin{definition}}
\def\ed{\end{definition}}
\def\bp{\begin{proposition}\rm}
\def\ep{\end{proposition}}
\def\bc{\begin{corollary}}
\def\ec{\end{corollary}}
\def\bl{\begin{lemma}\em}
\def\el{\end{lemma}}
\def\br{\begin{remark}\rm\small}
\def\er{\end{remark}}
\def\brs{\begin{remarks}.\\ \rm\
\begin{enumerate}}
\def\ers{\end{enumerate}\end{remarks}}

\def\ra{{\rightarrow}}

\def\Tr{\mathrm {Tr}}

\def\harm{\mathrm {harm}}
\def\tree{\mathrm {tree}}

\def\&{&{\hskip -20pt}}

\def\CC{{\mathcal C}}
\def\DD {{\mathcal D}}
\def\EE {{\mathcal E}}

\def\Rb{{\mathbf R}}

\date{}
\newcommand{\bibi}[5]{\bibitem{#1} \textsc{#2}, \textit{#3}, #4, #5}
\newcommand{\ds}{\ensuremath{\displaystyle}}

\newcommand{\order}[1]{\ensuremath{{\mathcal O}\left(#1\right)}}
\newcommand{\be}{\begin{equation}}
\newcommand{\ee}{\end{equation}}
\newcommand{\bea}{\begin{eqnarray}}
\newcommand{\eea}{\end{eqnarray}}
\newcommand{\supp}{\mathrm{supp}}

\renewcommand{\Re}{\mathrm{Re}}
\renewcommand{\Im}{\mathrm{Im}}
\newcommand{\res}{\textrm{Res}}
\newcommand{\R}{\ensuremath{\mathbb R}}
\newcommand{\C}{\ensuremath{\mathbb C}}
\newcommand{\N}{\ensuremath{\mathbb N}}
\newcommand{\M}{\ensuremath{\mathcal M}}

\renewcommand{\c}[1]{\ensuremath{\overline{#1}}}

\newtheorem{theorem}{Theorem}[section]
\newtheorem{definition}[theorem]{Definition}
\newtheorem{lemma}[theorem]{Lemma}
\newtheorem{proposition}[theorem]{Proposition}

\newtheorem{remark}{Remark}[section]

\newcounter{redboxes}
\newcommand{\semmi}[1]{}

\begin{document}
\baselineskip 16pt
\begin{flushright}
\end{flushright}
\medskip
\begin{center}
\begin{Large}\fontfamily{cmss}
\fontsize{17pt}{27pt} \selectfont \textbf{Superharmonic Perturbations of a Gaussian Measure, Equilibrium Measures and Orthogonal Polynomials}\footnote{ Work supported in part by the
Natural Sciences and Engineering Research Council of Canada
(NSERC) and the Fonds de recherche sur la nature et les technologies du Qu\'ebec (FQRNT) }
\end{Large}\\
\bigskip
\begin{large}  {F. Balogh}$^{\dagger \ddagger}$\footnote{harnad@crm.umontreal.ca}
 and {J. Harnad}$^{\dagger \ddagger}$\footnote{balogh@crm.umontreal.ca}
\end{large}
\\
\bigskip
\begin{small}
$^{\dagger}$ {\em Centre de recherches math\'ematiques,
Universit\'e de Montr\'eal\\ C.~P.~6128, succ. centre ville, Montr\'eal,
Qu\'ebec, Canada H3C 3J7} \\
\smallskip
$^{\ddagger}$ {\em Department of Mathematics and
Statistics, Concordia University\\ 7141 Sherbrooke W., Montr\'eal, Qu\'ebec,
Canada H4B 1R6} \\
\end{small}
\end{center}
\bigskip
\bigskip

\begin{center}{\bf Abstract}
\end{center}
\smallskip

\begin{small}

This work concerns superharmonic perturbations of a Gaussian measure given by a special class of positive weights in the complex plane of the form $w(z) = \exp(-|z|^2 + U^{\mu}(z))$,  where $U^{\mu}(z)$ is the logarithmic potential of a compactly supported positive measure $\mu$.  The equilibrium measure of the corresponding
weighted energy problem is shown to be supported on subharmonic generalized quadrature domains for a large class of perturbing potentials $U^{\mu}(z)$.  
It is also shown that the $2\times 2$ matrix d-bar problem for orthogonal polynomials with respect to such weights  is well-defined and has a unique solution given explicitly by  Cauchy transforms.
Numerical evidence is presented supporting a conjectured relation between the asymptotic distribution of the zeroes of the orthogonal polynomials in a semi-classical scaling limit and the Schwarz function of the curve bounding the support of the equilibrium measure, extending the previously studied case of harmonic polynomial perturbations with weights $w(z)$ supported on a compact domain.

\end{small}
\medskip




\section{Introduction}
This work mainly concerns equilibrium problems in potential theory, but its motivation derives largely from two related domains: random matrix theory and interface dynamics of incompressible fluids. 
Recent work of P. Wiegmann, A. Zabrodin and their collaborators  \cite{WZ, WZ1} connected the spectral distributions of random normal matrices to the \emph{Laplacian Growth} model for the interface dynamics of a pair of  two-dimensional incompressible fluids. The unitarily invariant probability measure on the set of $n\times n$ complex normal matrices in  \cite{WZ1} is determined by a potential function
\be
\label{matrixpotential}
V(z,\c{z}) = z\c{z} + A(z)+\c{A(z)}
\ee
and the corresponding density is of the form
\be
f(M, M^*)=e^{-\frac{1}{\hbar}\textrm{Tr}V(M,M^{*})}
\ee
with respect to a unitarily invariant reference measure on the set of normal matrices. The function $A(z)$ is assumed to have a single-valued derivative $A'(z)$  meromorphic in some domain $D \subset \C$. Under suitable assumptions on $A(z)$, the large $n$ limit of the averaged normalized eigenvalue distribution in the scaling limit
\be
\label{semiclassical}
n \to \infty, \quad \hbar \to 0, \quad n\hbar =t
\ee 
tends to a probability measure $\mu_{V,t}$ where $t$ is some fixed positive number (\emph{quantum area}). It turns out that $\mu_{V,t}$ is the unique solution of a two-dimensional electrostatic equilibrium problem in the presence of the external potential $V$ in the complex plane. In most cases, $\mu_{V,t}$ is absolutely continuous with respect to the planar Lebesgue measure with constant density. It can be shown that the support $\supp(\mu_{V,t})$ undergoes Laplacian growth in terms of the scaling constant $t$ and the area of the support is linear in $t$. However, this evolution problem is ill-defined; even initial domains with analytic boundaries may develop cusp-like singularities in finite time and the solution cannot be continued in the strong sense (see \cite{WZ1}, \cite{Teodorescu}). The averaged eigenvalue density of the corresponding normal matrix model for finite matrix size $n$ be viewed as describing a sort of discretized version whose contiuuum limit  may be interpreted as a semiclassical limit (\ref{semiclassical}) that tends to $\mu_{V,t}$ as shown in  \cite{HedenmalmMakarov, AmeurHedenmalmMakarov}.

In studying these questions, it is important to understand first the possible shapes of compact sets which are supports of equilibrium measures  for potentials of the form (\ref{matrixpotential}). In this work we show the support of the equilibrium measure  for a class of perturbed Gaussian potentials of the form
\be
V_{\alpha, \nu}(z) := \alpha|z|^2 + U^{\nu}(z),
\ee
where $\alpha >0$ and $\nu$ is a compactly supported finite positive Borel measure, to be  so-called \emph{generalized quadrature domains} \cite{GustafssonShapiro,  Sakai}.

To understand the asymptotic behaviour of the averaged eigenvalue density of normal matrix models in different scaling regimes, one has to study the asymptotics of the corresponding orthogonal polynomials for the weight $e^{-V(z)}$. The well-known Riemann-Hilbert approach is not applicable directly in this case because the orthogonality weight is not constrained to the real axis. However, there is a sort of matrix $\c{\partial}$-bar problem, introduced by Its and Takhtajan \cite{ItsTakhtajan}, which is a candidate to replace the matrix Riemann-Hilbert problem in the study of strong asymptotics of the orthogonal polynomials. In the present work in this regard is  shown that the $\c{\partial}$-bar problem is also well-defined and characterizes the orthogonal polynomials for the class of perturbed Gaussian potentials considered above.

To fix notations, let $m$ denote the two-dimensional Lebesgue measure in the complex plane $\C$. We denote  respectively by $\c{H}$ and $H^{c}$ the closure and the complement of a set $H \subset \C$
and by $ I_S$ the indicator function..
The open disk of radius $r$ centered at $c \in \C$ is denoted by $B(c,r)$, and 
the Riemann sphere by $\hat{\C}$.


\section{Weighted energy problem and logarithmic Potentials}
\setcounter{equation}{0}

In this section we briefly describe both the \emph{classical and weighted energy problems} of logarithmic potential theory 
(see \cite{Ransford} and \cite{SaffTotik}) and specify the class of background potentials we are concerned with in this paper.

\begin{definition}
Let $\mu$ be a compactly supported finite positive Borel measure in the complex plane. The \emph{logarithmic potential} produced by $\mu$
is defined as
\be
U^{\mu}(z) := \int_{\C}{\log{\frac{1}{|z-w|}}d\mu(w)} \qquad (z\in \C). 
\ee
\end{definition}
In particular, for a bounded subset $S$ of the plane with $m(S) >0$, we consider the measure $\eta_S$ given by
\be
d\eta_S = I_S dm.
\ee
Thus $\eta_S$ is the Lebesgue measure restricted to $S$. In the following, we use the simplified notation $U^{S}(z)$ for the logarithmic potential $U^{\eta_S}(z)$ of the measure $\eta_S$.

The logarithmic potential of a positive measure $\mu$ is harmonic outside the support of $\mu$ and superharmonic on $\supp(\mu)$ (see \cite{SaffTotik}, Theorem 0.5.6). Moreover, it has the asymptotic behaviour
\be
U^{\mu}(z) = \mu(\C)\log{\frac{1}{|z|}} +\order{\frac{1}{z}}, \qquad (|z| \to \infty)
\ee
where $\mu(\C)$ is the total mass of $\mu$.
If $U^{\mu}(z)$ is smooth enough the density of the measure $\mu$ can be recovered from this potential by taking the Laplacian of $U^{\mu}(z)$: 
\begin{theorem}[\cite{SaffTotik}, II.1.3]
\label{density}
If in a region $R \subseteq \C$ the logarithmic potential $U^{\mu}(z)$ of the measure $\mu$ has continuous second partial derivatives, then $\mu$ is absolutely continuous with respect to the planar Lebesgue measure $m$ in $R$ and we have the formula
\be
d\mu=-\frac{1}{2\pi}\Delta U^{\mu}dm.
\ee
\end{theorem}
Now, let $K$ be a compact subset of $\C$ and  let $\M(K)$ denote the set of all Borel probability measures supported on $K$.
In classical potential theory, the \emph{logarithmic energy} of a measure $\mu \in \M(K)$ is defined to be
\be
I(\mu) := \int_K{U^{\mu}(z)d\mu(z)} = \int_K\!\!\int_K{\log{\frac{1}{|z-t|}}d\mu(t)d\mu(z)}.
\ee
The quantity
\be
E_K := \ds \inf_{\mu \in \M(K)}{I(\mu)}
\ee
is either finite or $+\infty$. The \emph{logarithmic capacity} of $K$ is
\be
\textrm{cap}(K):= e^{-E_K}.
\ee
 If $E_K < \infty$ then, by a well-known theorem of Frostman (see e.g. \cite{Ransford}), there exists a unique measure $\mu_K$ in $\M(K)$ minimizing the energy functional $I(\cdot)$ and this measure is called the \emph{equilibrium measure} of $K$. The capacity of an arbitrary Borel set $B \subset \C$ is defined as
 \[ \textrm{cap}(B) := \sup\{\textrm{cap}(K)\ | \ K \subseteq B, K \textrm{ compact}\}.\]
 A property is said to hold \emph{quasi-everywhere} if the set of exceptional points (i.e. those where it does not hold) is of capacity zero.\\

In the more general setting we have a closed set $\Sigma \subseteq \C$ and a function $w\colon \Sigma \to [0,\infty)$ on $\Sigma$ called the \emph{weight function}. Usually the weight function is given in the form 
\be
w(z)=\exp(-Q(z))
\ee
where $Q \colon \Sigma \to (-\infty,\infty]$. In the electrostatical interpretation the set $\Sigma$ is called the \emph{conductor} and $Q$ is called the \emph{background potential}.
\begin{definition}[\cite{SaffTotik}]
\label{admissibility}
The weight function $w$ is said to be \emph{admissible} if 
\begin{itemize}
 \item $w$ is upper semi-continuous,
 \item $\{z \in \Sigma\ |\ w(z) >0 \}$ has nonzero capacity,
 \item $\ds \lim_{|z|\to \infty}|z|w(z)=0$.
\end{itemize}
\end{definition}
The admissibility conditions can be rephrased in terms of the potential $Q$; $w(z) =\exp(-Q(z))$ is admissible if and only if
$Q$ is lower semi-continuous, the set $\{z \in \Sigma\ |\ Q(z) < \infty\}$ has nonzero capacity and $\ds \lim_{|z|\to \infty}(Q(z)-\log|z|)=\infty$.

Let $\M(\Sigma)$ denote the set of all Borel probability measures supported on $\Sigma \subseteq \C$.
The \emph{weighted energy functional} $I_Q$ is defined for all $\mu \in \M(\Sigma)$ by
\bea
I_Q(\mu) &:=& \int_{\Sigma}\!\!\int_{\Sigma}{\log{\left[|z-t|w(z)w(t)\right]^{-1}}d\mu(z)d\mu(t)}\\
&=&\int_{\Sigma}\!\!\int_{\Sigma}{\log{\frac{1}{|z-t|}}d\mu(z)d\mu(t)} +2\int_{\Sigma}Q(z)d\mu(z).
\eea
The goal is then to find a probability measure that minimizes this functional on $\M(\Sigma)$.
If $Q$ is admissible it can be shown (see \cite{SaffTotik}, Theorem I.1.3) that
\be
 E_Q:=\inf_{\mu \in \M(\Sigma)}{I_Q(\mu)}
\ee
is finite and there exists a unique measure, denoted by $\mu_Q$, that has finite logarithmic energy and minimizes 
$I_Q$. Moreover, the support of $\mu_Q$, denoted by $S_Q$, is compact and has positive capacity. The measure $\mu_Q$ is called the \emph{equilibrium measure} of the background potential $Q$. The logarithmic potential satisfies the equilibrium conditions
\begin{eqnarray}
\label{eqcondition}
\nonumber
U^{\mu_Q}(z) +Q(z) &\geq& F_{Q} \textrm{ quasi-everywhere on }\Sigma,\\
U^{\mu_Q}(z) +Q(z) & \leq & F_{Q} \textrm{ for all } z \in S_Q,
\end{eqnarray}
where $F_Q$ is the \emph{modified Robin constant}:
\be
F_{Q} = E_Q -\int{Qd\mu_Q}.
\ee

Motivated by random normal matrix models (see \cite{WZ1}, \cite{Teodorescu}), we are interested in background potentials of the following type:
\begin{equation}
\label{background}
V_{\alpha, \nu}(z) := \alpha|z|^2 + U^{\nu}(z),
\end{equation}
where $\alpha$ is a positive real number and $\nu$ is a compactly supported finite positive Borel measure. These potentials have a planar Gaussian leading term controlled by the positive parameter $\alpha$ and this is perturbed by a fixed positive charge distribution given by the measure $\nu$. 

\begin{proposition} 
The potential $V_{\alpha, \nu}(z)$ is admissible for all possible choices of $\alpha$ and $\nu$.
\end{proposition}
\noindent
{\bf Proof.}
$V_{\alpha, \nu}(z)$ is lower semi-continuous because $U^{\nu}(z)$ is superharmonic in the whole complex plane. The set where $V_{\alpha, \nu}(z)$ is finite contains at least $\C\setminus \supp(\nu)$, which is of positive capacity since $\supp(\nu)$ is compact. Finally, the required boundary condition is also fulfilled:
\bea
V_{\alpha,\nu}(z) -\log|z| & = & \alpha|z|^2 +U^{\nu}(z)-\log|z| \\
& = & \alpha|z|^2 -\left(\nu(\C)+1\right)\log|z| +\order{\frac{1}{z}},
\eea
so the difference $V_{\alpha,\nu}(z) - \log|z|$ goes to $+\infty$ as $|z| \to \infty$.
\begin{flushright}
$\square$
\end{flushright}

We are especially interested in cases for which the perturbing measure $\nu$ is singular with respect to the planar Lebesgue measure $m$.
In particular, $\nu$ can be chosen to be a positive linear combination of point masses, i.e.
\be
\nu := \sum_{k=1}^{m}\beta_k \delta_{a_k},\quad \beta_k \in \R^{+}
\ee
where $a_1, a_2, \dots, a_m \in \C$ are the locations and $\beta_1, \beta_2, \dots, \beta_m$ are the charges of the fixed point masses respectively.


\section{Supports of equilibrium measures and quadrature domains}
\setcounter{equation}{0}

The determination of the support of the equilibrium measure for a background potential $V(z)=V_{\alpha,\nu}(z)$ of the form (\ref{background}) above is closely related to finding  \emph{generalized quadrature domains} of some measures in the complex plane. Let us recall the definition of quadrature domains given by M. Sakai (\cite{Sakai}).
\begin{definition}
Let $\nu$ be a positive Borel measure on the complex plane.
For a nonempty domain (open, connected) $\Omega$ in $\C$
let $F(\Omega)$ be a subset of the space
\be
\Re L^1(\Omega) := \left\{\Re{f} \ |\ f \in L^1(\Omega, m)\right\}.
\ee
of real-valued integrable functions on $\Omega$.

The domain $\Omega$ is called a \emph{(generalized) quadrature domain} of the measure $\nu$ for the function class $F(\Omega)$ if
\begin{enumerate}
\item $\nu$ is concentrated on $\Omega$, i.e. $\nu\left(\Omega^c\right)=0$,
\item
\be
\int_{\Omega}{f^{+}d\nu} < \infty \quad \textrm{\it and } \quad \int_{\Omega}f d\nu \leq \int_{\Omega}f dm
\ee
for every $f \in F(\Omega)$ where $f^{+} := \max\{f,0\}$.
\end{enumerate}
\end{definition}
Note that if $F(\Omega)$ is a function class such that $-f \in F(\Omega)$ whenever $f \in F(\Omega)$ then 
the second condition is equivalent to
\be
\int_{\Omega}{|f|d\nu} < \infty \quad \textrm{ and } \quad \int_{\Omega}f d\nu = \int_{\Omega}f dm
\ee
for every $f \in F(\Omega)$. We are interested in the following subclasses:
\bea
\Re AL^1(\Omega) &=& \{\Re{f} \in L^1(\Omega,m) \ |\ f \textrm{ is holomorphic in }\Omega \} \\
HL^1(\Omega) &=& \{h \in L^1(\Omega,m) \ |\ h \textrm{ is harmonic in }\Omega \} \\
SL^1(\Omega) &=& \{s \in L^1(\Omega,m) \ |\ s \textrm{ is subharmonic in }\Omega \}
\eea
For a measure $\nu$ the quadrature domains corresponding to these classes are called classical (holomorphic), harmonic and subharmonic quadrature domains respectively. We have the obvious inclusions
\be
Q(\nu,SL^1) \subseteq Q(\nu,HL^1) \subseteq Q(\nu,\Re AL^1),
\ee
where  $Q(\nu, F)$ denotes the set of quadrature domains of $\nu$ for the function class $F$. 
It is important to note that if the domain $\Omega$ belongs to $Q(\nu, F)$ then its \emph{saturated set} or \emph{areal maximal set}
\be
[\Omega] := \left\{z \in \C \ |\ m\left(B(z,r)\cap \Omega ^c\right) =0 \textrm{ for some } r >0 \right\}
\ee
also belongs to $Q(\nu,F)$.

For example, it can be shown that the disk $B(c,R)$ is the only classical quadrature domain for the point measure $\nu = R^2 \pi\delta_{c}$ (see \cite{Sakai}, Example 1.1). The simplest examples are the \emph{classical quadrature domains} whose quadrature measure is a positive linear combination of point masses:
\be
\nu = \sum_{k=1}^{m}\beta_k \delta_{a_k},\quad \beta_k \in \R^{+}.
\ee
This means that for every holomorphic function $f$ that is integrable on $\Omega$ we have the identity
\be
\int_{\Omega}f dm = \sum_{k=1}^{m}\beta_k f(a_k).
\ee
An immediate generalization is obtained by allowing points $a_k$ of higher multiplicities $j_k \geq 1$ in the above sum, this means allowing derivatives of finite order to appear in the sum representing the area integral functional:
\be
\int_{\Omega}f dm = \sum_{k=1}^{m}\sum_{l=0}^{j_k}\beta_{k,l} f^{(l)}(a_k).
\ee
However, in this work we do not consider such quadrature domains.

It is easy to see that if $\Omega$ is a subharmonic quadrature domain of the measure $\nu$ then, using subharmonic test functions of the form 
\be
s_{z}(w) = -\log{\frac{1}{|z-w|}} \qquad (z \in \C),
\ee
we have
\bea
U^{\Omega}(z) & \leq & U^{\nu}(z) \textrm{ if } z \in \C,\cr
U^{\Omega}(z) & = & U^{\nu}(z) \textrm{ if } z \in \C\setminus \Omega.
\eea
To illustrate the structure of the equilibrium measure of a potential of the form (\ref{background}) above, we consider a simple but nontrivial example:
\be
V(z) = \alpha|z|^2 +\beta\log{\frac{1}{|z-a|}},
\ee 
where $\alpha \in \R^{+}, \beta \in \R^{+}$ and $a \in \C$. The calculation of the equilibrium measure for this potential is quite standard (see, for example, \cite{WZ1}, \cite{Teodorescu}) but the details will be of importance in suggesting generalizations. For the sake of completeness, the statement of the result and a short sketch of its proof are therefore included here.

\begin{figure}[htb]
\begin{center}
\begin{minipage}{0.49\textwidth}
\begin{center}
\includegraphics[width=5cm]{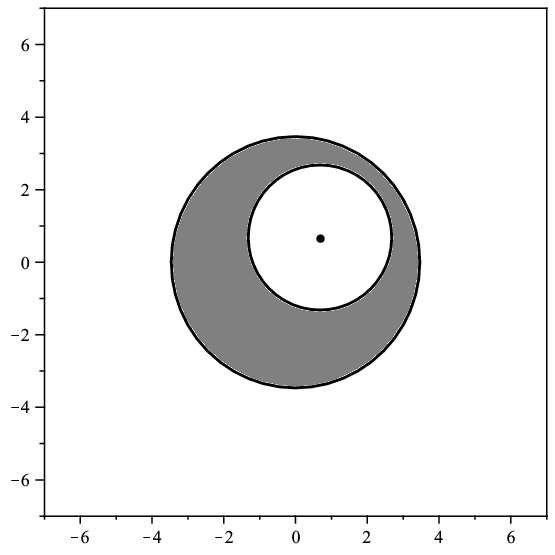}\\
(a.1) 
\end{center}
\end{minipage}
\begin{minipage}{0.49\textwidth}
\begin{center}
\includegraphics[width=5cm]{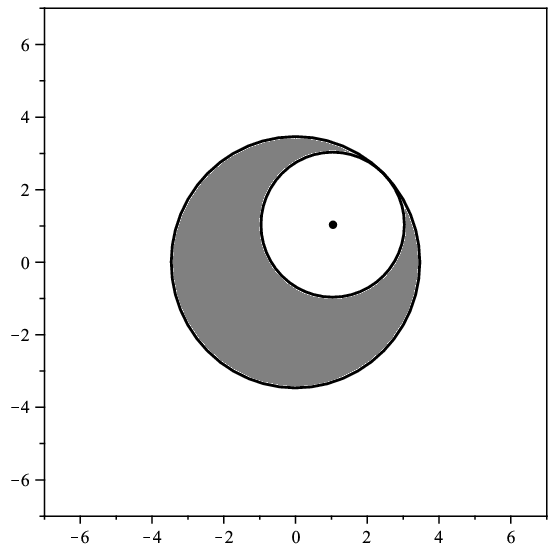}\\
(a.2) 
\end{center}
\end{minipage}
\vskip 0.3cm
\begin{minipage}{0.49\textwidth}
\begin{center}
\includegraphics[width=5cm]{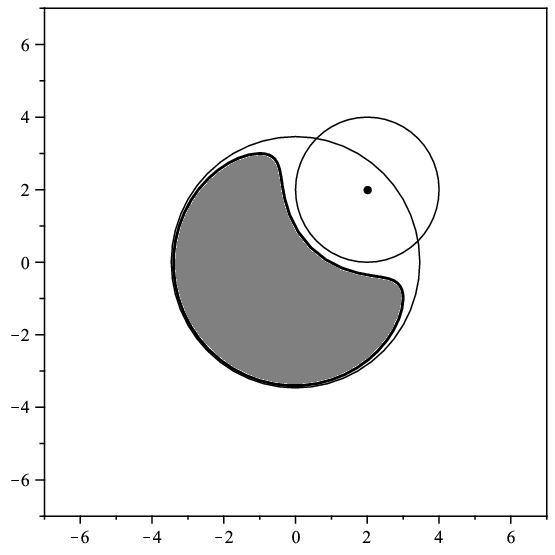}\\
(b.1)
\end{center}
\end{minipage}
\begin{minipage}{0.49\textwidth}
\begin{center}
\includegraphics[width=5cm]{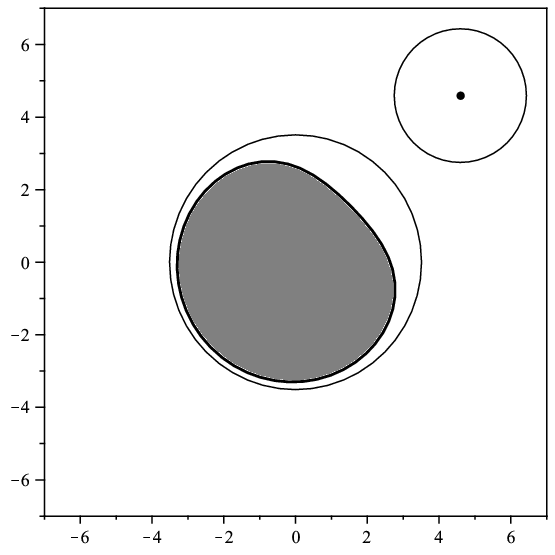}\\
(b.2)
\end{center}
\end{minipage}
\caption{The shape of the support $S_V$ (shaded area) is illustrated for the different disk configurations (a) and (b) in Proposition \ref{onepoint}.}
\label{pcfigures}
\end{center}
\end{figure}

To find the equilibrium measure for this potential one can use the following characterization theorem:
\begin{theorem}[\cite{SaffTotik}, I.3.3]
\label{characterization}
Let $Q \colon \Sigma \to (-\infty, \infty]$ be an admissible background potential. If a measure $\sigma \in \M(\Sigma)$ has compact support and finite logarithmic energy, and there is a constant $F \in \R$ such that
\begin{equation}
\label{eqcond1}
U^{\sigma}(z)+Q(z) = F \ \textrm{ quasi-everywhere on } \textrm{supp}(\sigma)
\end{equation}
and 
\begin{equation}
\label{eqcond2}
U^{\sigma}(z)+Q(z) \geq F \ \textrm{ quasi-everywhere on } \Sigma,
\end{equation}
then $\sigma$ coincides with the equilibrium measure $\mu_Q$.
\end{theorem}
The logarithmic potential of the uniform measure $\eta_{B(c,R)}$ on a disk $B(c,R)$ is easily calculated to be
\begin{equation}
\label{logpotential}
U^{B(c,R)}(z)=
\left\{
\begin{array}{cc}
\ds \frac{1}{2}R^2\pi\left(\log{\frac{1}{R^2}}+1-\frac{|z-c|^2}{R^2}\right) & |z-c| \leq R \\
& \\
\ds R^2\pi\log{\frac{1}{|z-c|}} & |z-c| > R.
\end{array}
\right.
\end{equation}
\begin{proposition}
\label{onepoint}
Define two radii $R$ and $r$ as 
\be
R  := \sqrt{\frac{1+\beta}{2\alpha}} \quad \textrm{\it and } \quad r  :=  \sqrt{\frac{\beta}{2\alpha}}.
\ee
The equilibrium measure $\mu_V$ is absolutely continuous with respect to the Lebesgue measure and its density is the constant $\frac{2\alpha}{\pi}$. The support $S_V$ of $\mu_V$ depends on the geometric arrangement of the disks $B(a,r)$ and $B(0,R)$ in the following way:
\begin{enumerate}
\item If $B(a,r) \subset B(0,R)$ then
\be
S_{V} = \c{B(0,R)}\setminus B(a,r)
\ee
{\rm (see (a.1) and (a.2) in Figure \ref{pcfigures})}.
\item If $B(a,r) \not \subset B(0,R)$ then $\hat{\C} \setminus S_V$ is given by a rational exterior conformal mapping of the form
\be
f \colon \hat{\C} \setminus \{ \zeta \ \colon \ |\zeta|  \leq 1 \} \to \hat{\C} \setminus S_V, \qquad f(\zeta) = \rho\zeta +u +\frac{v}{\zeta - A},
\ee 
where the coefficients $\rho \in \R^{+}, 0 < |A| < 1$ and $u,v \in \C$ of the mapping $f(\zeta)$ are uniquely determined by the parameters $\alpha, \beta$ and $a$ of the potential $V(z)$
{\rm (see (b.1) and (b.2) in Figure \ref{pcfigures})}.
\end{enumerate}
\end{proposition}
\noindent
{\bf Proof.} Suppose first that $B(a,r) \subset B(0,R)$.\\
Let $\sigma$ be the measure given by
\be 
\ds d\sigma := \frac{1}{m(K)}I_K dm 
\ee
where $K = \c{B(0,R)}\setminus B(a,r)$. The area of $K$ is 
\be
\label{areacond}
m(K)=(R^2-r^2)\pi = \frac{\pi}{2\alpha}.
\ee
Therefore the logarithmic potential of $\sigma$ is
\be
U^{\sigma}(z) = \frac{2\alpha}{\pi}\left(U^{B(0,R)}(z)-U^{B(a,r)}(z)\right).
\ee
Now, for $z \in K$ the effective potential at $z$ is
\bea
U^{\sigma}(z) + V(z) & = & \alpha R^2\left(\log{\frac{1}{R^2}}+1-\frac{|z|^2}{R^2}\right) - 2\alpha r^2\log{\frac{1}{|z-a|}}\cr
&  & + \alpha|z|^2 +\beta\log{\frac{1}{|z-a|}}\cr
& = & \alpha R^2\left(\log{\frac{1}{R^2}}+1\right).
\eea
Define 
\be
F := \alpha R^2\left(\log{\frac{1}{R^2}}+1\right).
\ee
If $z \not \in K$ a short calculation gives
\be
U^{\sigma}(z) + V(z) =
\left\{
\begin{array}{ll}
\ds F +2\alpha r^2 f\left(\frac{|z-a|}{r} \right) & z \in B(a,r) \\
\ds F + 2\alpha R^2 f\left( \frac{|z|}{R}\right) & |z| > R.
\end{array}
\right.
\ee
where 
\be
f(x) := \frac{x^2-1}{2}-\log{x}.
\ee
Since $f$ is nonnegative the effective potential satisfies
\be
U^{\sigma}(z) + V(z) \geq  F.
\ee
By Theorem \ref{characterization}, we conclude that the equilibrium measure for the background potential $V$ is $\sigma$.

Now suppose that $B(a,r) \not \subset B(0,R)$. For this case, we only sketch the calculation giving the system of equations that relates the parameters of the potential $V(z)$ and the conformal map $f(\zeta)$.
The potential is smooth in  the domain $\C\setminus \{a\}$ and its Laplacian there is constant which suggests, by Theorem \ref{density}, that the density of the equilibrium measure is equal to
\be
\frac{1}{2\pi}\Delta V(z) = \frac{2\alpha}{\pi}.
\ee
We expect the equilibrium measure therefore to be the normalized Lebesgue measure restricted to some compact set $S \subset \C \setminus \{a\}$; that is,
\be
d\mu_V = \frac{1}{m(S)}dm
\ee
where $m(S)$ is given by (\ref{areacond}).
The first equilibrium condition (\ref{eqcond1}) is
\be
U^{\mu_V}(z) + V(z) = F \qquad (z \in S)
\ee
for some constant $F$. Assuming the necessary smoothness of $U^{\mu_V}(z)$ and applying the differential operator $\partial_{z}$, this gives
the necessary condition
\be
-\frac{\alpha}{\pi}\int_{S}{\frac{dm(w)}{z-w}} +  \left(\alpha\c{z} -\frac{\beta}{2}\frac{1}{z-a}\right)= 0 \qquad (z \in S),
\ee
which is the same as
\be
\frac{\alpha}{2\pi i}\int_{S}{\frac{d\c{w}\wedge dw}{z-w}} -   \alpha\c{z} +\frac{\beta}{2}\frac{1}{z-a}= 0 \qquad (z \in S).
\ee
By Green's Theorem we get
\be
\frac{1}{2\pi i}\int_{\partial S}{\frac{\c{w} dw}{z-w}} = -\frac{\beta}{2\alpha}\frac{1}{z-a} \qquad (z \in S).
\ee
Following \cite{WZ1}, we seek to express the compact set $S$ in terms of an exterior conformal mapping of the form
\be
f \colon \hat{\C} \setminus \{ \zeta \ \colon \ |\zeta|  \leq 1 \} \to \hat{\C} \setminus S, \qquad f(\zeta) = \rho\zeta +u+\frac{v}{\zeta - A}
\ee 
where $\rho > 0$ and $0 < |A| <1$. By rewriting the previous form of the derived equilibrium condition we obtain
\be
\frac{1}{2\pi i}\int_{|\zeta|=1}{\frac{\c{f(\zeta)}f'(\zeta) d\zeta}{z-f(\zeta)}} = -\frac{\beta}{2\alpha}\frac{1}{z-a} \qquad (z \in S).
\ee
Along the positively oriented simple circular contour $|\zeta|=1$ we have
\be
\c{f(\zeta)} = \rho\frac{1}{\zeta} +\c{u}+\frac{\c{v}\zeta}{1 - \c{A}\zeta}.
\ee
Therefore the rational function
\begin{equation}
\label{T}
T(\zeta;z) := \frac{\left(\rho\frac{1}{\zeta} +\c{u}+\frac{\c{v}\zeta}{1 - \c{A}\zeta}\right)\left(\rho -\frac{v}{(\zeta - A)^2}\right)}
{z-\rho\zeta -u-\frac{v}{\zeta - A}}
\end{equation}
has coefficients depending on $\rho, u, v, A$ and satisfies the equation
\be
\frac{1}{2\pi i}\int_{|\zeta|=1}{T(\zeta;z)d\zeta} = -\frac{\beta}{2\alpha}\frac{1}{z-a} \qquad (z \in S)
\ee
subject to the area normalization condition (\ref{areacond}).

The differential $T(\zeta; z)d\zeta$ has four fixed poles in $\zeta$ at $\zeta = 0$, $\zeta=\infty$, $\zeta = A$, $\zeta = \frac{1}{\c{A}}$ and two other poles  depending on $z$ via the equation $z =f(\zeta)$. (These extra poles may coincide with some of the fixed poles above.) This equation can be rewritten as a quadratic equation in $\zeta$:
\be
\rho\zeta^2+(u-z-A\rho)\zeta+A(z-u)+v = 0.
\ee
If $z \in S$ both solutions of this equation are \emph{inside} the unit disk $\{\zeta \ |\ |\zeta| < 1\}$. To calculate the integral of $T(\zeta;z)d\zeta$ in terms of residues, we write the the contour integral in the standard local coordinate $\xi =\frac{1}{\zeta}$ around $\zeta=\infty$:
\be
\frac{1}{2\pi i}\int_{|\zeta|=1}{T(\zeta;z)d\zeta} = \frac{1}{2\pi i}\int_{|\xi|=1}{T\left(\frac{1}{\xi};z\right)\frac{d\xi}{\xi^2}}.
\ee
The two simple poles inside the disk $\{\xi \ |\ |\xi| < 1\}$ are $\xi=0$ and $\xi = \c{A}$ and hence
\be
\frac{1}{2\pi i}\int_{|\xi|=1}{T\left(\frac{1}{\xi};z\right)\frac{d\xi}{\xi^2}} = \res_{\xi=0}{T\left(\frac{1}{\xi};z\right)\frac{1}{\xi^2}} +
\res_{\xi=\c{A}}{T\left(\frac{1}{\xi};z\right)\frac{1}{\xi^2}}.
\ee
Since
\be
T\left(\frac{1}{\xi};z\right)\frac{1}{\xi^2} = 
\ds \frac{1}{\xi}\frac{\left(\rho\xi +\c{u}+\frac{\c{v}}{\xi - \c{A}}\right)\left(\rho -\frac{v\xi^2}{(1 - A\xi)^2}\right)(1-A\xi)}
{(1-A\xi)(z\xi-u\xi-\rho)-v\xi^2},
\ee
the residues are
\bea
\res_{\xi=0}{T\left(\frac{1}{\xi};z\right)\frac{1}{\xi^2}} & = & \frac{\c{v}}{\c{A}}-{\c{u}},\\
\res_{\xi=\c{A}}{T\left(\frac{1}{\xi};z\right)\frac{1}{\xi^2}} & = & \frac{\c{v}}{\c{A}}\frac{\left(\rho -\frac{v\c{A}^2}{(1 - |A|^2)^2}\right)(1-|A|^2)}
{(1-|A|^2)(z\c{A}-u\c{A}-\rho)-v\c{A}^2}.
\eea
A short calculation gives the area of $S$ in terms of the mapping parameters:
\be
m(S) = \pi\left(\rho^2 -\frac{|v|^2}{(1-|A|^2)^2}\right).
\ee
Finally we obtain the following system of equations:
\be
\label{system}
\begin{array}{ccc}
\ds \rho^2 -\frac{|v|^2}{(1-|A|^2)^2} & = & \ds \frac{1}{2\alpha} \\
\ds \frac{\c{v}}{\c{A}}-{\c{u}} & = & \ds 0 \\
\ds u + \frac{\rho}{\c{A}} + \frac{v\c{A}}{1-|A|^2} & = & \ds a\\
\ds \frac{\c{v}}{\c{A}^2}\left(\rho - \frac{v\c{A}^2}{(1-|A|^2)^2}\right) &= & \ds -\frac{\beta}{2\alpha}. 
\end{array}
\ee
We must prove that if we assume $|a| + r > R$ there exists a unique solution $\rho, u,v, A$ to this system in terms of the parameters $\alpha, \beta, a$. Eliminating $u$ from the third equation gives
\be
\label{phases}
a = \frac{\rho}{\c{A}} + \frac{v}{A(1-|A|^2)}.
\ee
The last equation of (\ref{system}) shows that $\frac{v}{A^2}$ is real. 
The phases of $v$ and $A$ are therefore fixed by (\ref{phases}). Writing $a$ in polar form
\be
a = t e^{i\phi}
\ee
(with $t >0$ because $B(a,r) \not \subset B(0, R)$), we obtain
\be
\quad v = s e^{2i\phi}, \quad A = K e^{i\phi},
\ee
where $s$ and $K$ are positive real numbers. We can express $\rho$ and $s$ in terms of $K$:
\bea
\rho & =& \frac{1}{2Kt}\left(K^2t^2 +\frac{1}{2\alpha}\right)\\
s &=& \frac{1-K^2}{2Kt}\left(K^2t^2 -\frac{1}{2\alpha}\right),
\eea
Setting $x=K^2$, this must be a solution of the the cubic equation
\be
\label{cubic}
2t^4 x^3 -\left(t^4 +\frac{1+2\beta}{\alpha}t^2 \right) x^2 +\frac{1}{4\alpha^2}=0.
\ee
The condition $|a| +r > R$ means that
\be
\label{kilog}
t^4 -\frac{1+2\beta}{\alpha} t^2 +\frac{1}{4\alpha^2} < 0.
\ee
Defining the function
\be
g(x) :=2t^4 x^3 -\left(t^4 +\frac{1+2\beta}{\alpha}t^2 \right) x^2 +\frac{1}{4\alpha^2},
\ee
we have
\be
g(0) = \frac{1}{4\alpha^2} >0 \ \textrm{ and }\ g(1) = t^4 -\frac{1+2\beta}{\alpha}t^2 +\frac{1}{4\alpha^2} < 0
\ee
by (\ref{kilog}). Since 
\bea
g'(x) &=&  6t^4 x^2 -2\left(t^4 +\frac{1+2\beta}{\alpha}t^2 \right) x \cr
&=& 6t^4 x \left(x-\frac{1}{3t^4}\left(t^4 +\frac{1+2\beta}{\alpha}t^2  \right)\right)
\eea
is negative in the interval $[0,1]$, $g(x)$ has a unique root in $(0,1)$, and herefore $K$ is uniquely determined by $(\ref{cubic})$.
This means that there is a unique solution for $\rho, u, v$ and $A$ of (\ref{system}) in terms of $\alpha, \beta, a$.

To conclude the proof one should show that the logarithmic potential $\frac{2\alpha}{\pi}U^{S}(z)$ satisfies the inequality (\ref{eqcond2}) of Theorem \ref{characterization}. This part of the proof is omitted.
\begin{flushright}
$\square$
\end{flushright}

\begin{figure}[htb]
\begin{center}
\includegraphics[width=6cm]{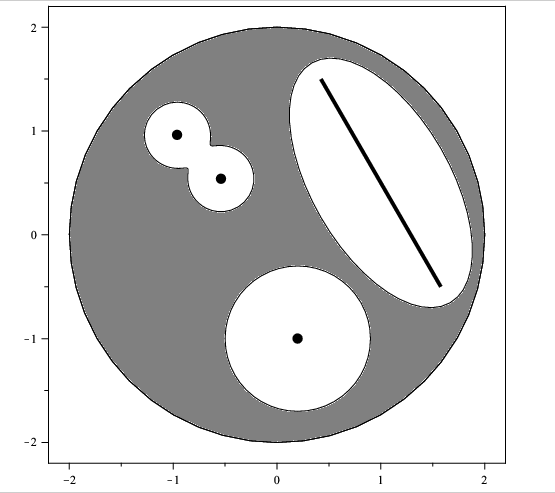}
\caption{A typical configuration involving subharmonic quadrature domains: a disk, a so-called \emph{bicircular quartic} (see \cite{Shapiro}) and an ellipse}
\label{melange}
\end{center}
\end{figure}

The electrostatic interpretation of case (a) in Proposition \ref{onepoint} is simple. If we replace the point charge $\beta\delta_{a}$ by a uniform charge distribution of density $\frac{2\alpha}{\pi}$ on $B(a,r)$, the resulting configuration is in equilibrium in the presence of the pure Gaussian potential $\alpha|z|^2$. The disk $B(a,r)$ is a quadrature domain for the measure $r^2\pi \delta_{a}$, so 
\be
\frac{2\alpha}{\pi}U^{B(a,r)}(z) = \beta U^{\delta_{a}}(z) \quad \textrm{ in } z\in B(a,r)^{c},
\ee
which means that the electric fields of $\frac{2\alpha}{\pi}\eta_{B(a,r)}$ and $\beta\delta_{a}$ are indistinguishable in the exterior of $B(a,r)$. A quadrature domain shaped cavity emerges in the support of the equilibrium measure of the unperturbed Gaussian potential since the fixed perturbing measure substitutes the portion of uniform charge placed in the cavity of the original equilibrium configuration, as illustrated in (a.1) and (a.2) of Figure \ref{pcfigures}.
A useful generalization of this idea turns out to be valid in a more general setting:

\begin{theorem}
\label{qds}
Let 
\be
V(z) := \alpha|z|^2 + U^{\nu}(z)
\ee
be a background potential. Assume that the measure $\nu$ can be decomposed into a sum
\be
\nu = \sum_{k=1}^{m}\nu_k
\ee
where the measures $\nu_k$ are all finite positive Borel measures satisfying the following conditions:
\begin{enumerate}
\item The supports of the measures $\nu_k$ are pairwise disjoint and each $\nu_k$ has positive total mass.
\item The measure $\frac{\pi}{2\alpha}\nu_k$ has an essentially unique (i.e. unique up to sets of measure zero) subharmonic quadrature domain, and $D_k$ denotes the saturated element of $Q(\nu_k,SL^{1})$ for all $k=1,2,\dots, m$ respectively.
\item The domains $D_k$ are pairwise disjoint and $D_k \subset B(0, R)$ for all $k=1,2,\dots,m$ where
\be
R:= \sqrt{\frac{1+\nu(\C)}{2\alpha}}.
\ee
\end{enumerate}
Then the equilibrium measure $\mu_V$ is absolutely continuous with respect to the Lebesgue measure with constant density $\frac{2\alpha}{\pi}$ and is supported on the set
\be
K:= \c{B(0,R)}\setminus \left(\bigcup_{k=1}^{m}D_k\right).
\ee
\end{theorem}
\noindent
(The situation is illustrated for a simple configuration of point and line charges in Figure \ref{melange}.)

\noindent
{\bf Proof.} Let $\sigma$ be the measure given by
\be 
\ds d\sigma := \frac{1}{m(K)}I_K dm 
\ee
To calculate the area of $K$, we note that the area of $D_k$ is given by
\be
m(D_k)  =  \int_{D_k}dm = \frac{\pi}{2\alpha} \int  d\nu_k = \frac{\pi}{2\alpha}\nu_k(\C).
\ee
Therefore 
\bea
m(K) &=& m(B(0,R)) - \sum_{k=1}^{m}m(D_k) \\
& = & \frac{\pi}{2\alpha}\left(1+\nu(\C) - \sum_{k=1}^{m}\nu_k(\C)\right)\\
& = & \frac{\pi}{2\alpha}.
\eea
For each measure $\nu_k$ the corresponding logarithmic potential $U^{\nu_k}(z)$ satisfies
\bea
\frac{2\alpha}{\pi}U^{D_k}(z) & \leq & U^{\nu_k}(z) \textrm{ if } z \in \C,\\
\frac{2\alpha}{\pi}U^{D_k}(z) & = & U^{\nu_k}(z) \textrm{ if } z \in \C\setminus D_k.
\eea
The logarithmic potential of $\sigma$ is
\be
U^{\sigma}(z) = \frac{2\alpha}{\pi}\left(U^{B(0,R)}(z)-\sum_{k=1}^{m}U^{D_k}(z)\right).
\ee
Now, if $z \in K$ then the effective potential at $z$ is
\bea
U^{\sigma}(z) + V(z) & = & \alpha R^2\left(\log{\frac{1}{R^2}}+1-\frac{|z|^2}{R^2}\right) - \frac{2\alpha}{\pi}\sum_{k=1}^{m}U^{D_k}(z)\cr
&  & + \alpha|z|^2 +U^{\nu}(z)\cr
& = & \alpha R^2\left(\log{\frac{1}{R^2}}+1\right) - \sum_{k=1}^{m}U^{\nu_k}(z) +U^{\nu}(z)\cr
& = & \alpha R^2\left(\log{\frac{1}{R^2}}+1\right).
\eea
Define $F := \alpha R^2\left(\log{\frac{1}{R^2}}+1\right)$.
If $z \not \in K$ then either $z \in D_k$ for some $k$ or $|z| >R$. If $z \in D_k$ then we have the inequality
\bea
U^{\sigma}(z) + V(z) & = & \alpha R^2\left(\log{\frac{1}{R^2}}+1-\frac{|z|^2}{R^2}\right) - \frac{2\alpha}{\pi}\sum_{k=1}^{m}U^{D_k}(z)\cr
&  & + \alpha|z|^2 +U^{\nu}(z)\cr
& = & F  + U^{\nu_k}(z) -\frac{2\alpha}{\pi}U^{D_k}(z)\cr
&   \geq & F.
\eea
On the other hand, if $|z| > R$ then
\bea
U^{\sigma}(z) + V(z) & = & 2\alpha R^2\log{\frac{1}{|z|}} - \frac{2\alpha}{\pi}\sum_{k=1}^{m}U^{D_k}(z)\cr
&  & + \alpha|z|^2 +U^{\nu}(z)\cr
& = & \alpha|z|^2 + \alpha R^2\log{\frac{1}{|z|^2}}\cr
& = & F  + 2\alpha R^2 f\left(\frac{|z|}{R}\right)\cr
&  \geq & F,
\eea
where $f(x) = \frac{x^2-1}{2}-\log{x}$. Since $f$ is nonnegative the effective potential satisfies
\be
U^{\sigma}(z) + V(z) \geq  F.
\ee
By Theorem \ref{characterization}, we conclude that the equilibrium measure for the background potential $V$ is $\sigma$.
\begin{flushright}
$\square$
\end{flushright}

The conclusion of Theorem \ref{qds} does not hold if some of the domains $D_k$ overlap or intersect the exterior of $B(0,R)$. 
In the first case we have to find a new decomposition of the perturbing measure and the corresponding domains; in the second the outer boundary no longer coincides with the boundary of $B(0,R)$ as we saw in Prop. \ref{onepoint}. It is hard to give a complete description of the support of the equilibrium measure in the general case. If $\nu$ is a finite linear combination of point masses then the methods of D. Crowdy and J. Marshall in \cite{CrowdyMarshall} used in the fluid dynamical context of rotating vortex patches are applicable to recover the corresponding supports.

In the cases considered above, the support $S_V$ of the equilibrium measure is contained the closed disk $\c{B(0,R)}$. It seems plausible that the same is true for all perturbing measures $\nu$. The following theorem states that $S_V \subset \c{B(0,R)}$ is valid if $\nu$ is a positive rational linear combination of point masses.

\begin{theorem}
Let $V(z)$ be a potential of the form
\be
V(z) = \alpha|z|^2 + U^{\nu}(z)
\ee
where $\nu$ is a measure of the form
\be
\nu = \sum_{k=1}^{m}{r_k \delta_{a_k}}
\ee
where $r_1, r_2, \dots, r_m$ are positive \emph{rational numbers}.. Then the support $S_V$ of the corresponding equilibrium measure is contained in the closed disk $\c{B(0,R)}$ where
\be
R= \sqrt{\frac{1+\nu(\C)}{2\alpha}}.
\ee
\end{theorem}

\noindent
{\bf Proof.} For a continuous weight $w(z)=\exp(-V(z))$ in the complex plane, $z \in \C$ belongs to the support $S_V$ if and only if for every neighborhood $B$ of $z$ there exists a \emph{weighted polynomial} $w^nP_n$ of degree $\deg{P_n}\leq n$, such that $w^nP_n$ attains its maximum modulus only in $B$ (see \cite{SaffTotik}, Corollary IV.1.4). Since our $w(z)$ is continuous this characterization is applicable to this setting.

Let $z \in S_V$ and suppose $B$ is a neighborhood of $z$. Then there exists a polynomial of degree at most $n$ for some $n \in \N$ such that $P_n(z)w^n(z)$ attains its maximum modulus only in $B$. Let $q$ be the least common denominator of the rational numbers $r_1, r_2, \dots, r_n$ such that
\be
r_k = \frac{p_k}{q}
\ee
where $p_k \in \N$ for all $k=1,2,\dots, m$. Then
\bea
\left(|P_n(z)|w^n(z)\right)^{q} & = & \left|P_n(z)^{q}\right|\left|\prod_{k=1}^{m}(z-a_k)^{n p_k}\right| e^{-nq\alpha|z|^2}.\cr
& = & \left|P_n(z)^q\prod_{k=1}^{m}(z-a_k)^{n p_k}\right| \exp\left(-n(q+L)\frac{q\alpha}{q+L}|z|^2\right),
\eea
where 
\be 
L = \sum_{k=1}^{m} p_k.
\ee 
Since all the $p_k$'s are assumed to be positive integers,
\be
Q_{n(q +L)}(z) := P_n(z)^{q}\prod_{k=1}^{m}(z-a_k)^{n p_k}
\ee
is a polynomial of degree at most $n(q +L)$. If we consider the modified weight $v(z) = \exp\left(-\frac{q\alpha}{q+L}|z|^2\right)$ the corresponding weighted polynomial 
\be
Q_{n(q+L)}(z)v^{n(q+L)}(z)
\ee
 attains its maximum modulus only in $B$. Therefore $z$ belongs to the support of the equilibrium measure of the weight $v(z)$ which is exactly $\c{B(0, R)}$ where
\be
R = \sqrt{\frac{q+L}{2q\alpha}} = \sqrt{\frac{1+\nu(\C)}{2\alpha}}.
\ee
This proves that $S_V \subseteq \c{B(0,R)}$.
\begin{flushright}
$\square$
\end{flushright}


\section{Orthogonal polynomials}
\setcounter{equation}{0}

In random normal matrix models, the correlation functions are expressed in terms of planar orthogonal polynomials with respect to \emph{scaled weight functions} of the form $\exp(-NQ(z))$ associated to a potential $Q(z)$ where $N > 0$ is a scaling parameter ($N$ has the same role as $\frac{1}{\hbar}$).
For our special potentials of the form $V(z) =V_{\alpha, \nu}(z)$ defined in (\ref{background}) above we have the weights
\be
e^{-NV(z)} = \exp\left(-N\left[\alpha|z|^2 + \int{\log{\frac{1}{|z-w|}}d\nu(w)}\right]\right),
\ee
where $N >0$ is the scaling parameter.
\begin{proposition}
\label{l1linfty}
We have 
\be 
e^{-NV(z)} \in L^1(\C,dm) \cap L^{\infty}(\C,dm)
\ee
for all choices of the parameters $N, \alpha, \nu$. Moreover, the absolute moments
\be
\int_{\C}|z|^k e^{-NV(z)}dm(z)
\ee
are all finite for $k =0,1,\dots$
\end{proposition}
\noindent
{\bf Proof.}
The exponent in the weight can be decomposed as
\be
N\left[\alpha|z|^2 + \int{\log{\frac{1}{|z-w|}}d\nu(w)}\right] = \frac{N\alpha}{2}|z|^2 + N\left[\frac{\alpha}{2}|z|^2+\int{\log{\frac{1}{|z-w|}}d\nu(w)}\right],
\ee
in which the second term is lower semicontinuous and satisfies
\be
N\left[\frac{\alpha}{2}|z|^2+\int{\log{\frac{1}{|z-w|}}d\nu(w)}\right]=N\left[\frac{\alpha}{2}|z|^2 +\nu(\C)\log{\frac{1}{|z|}}\right] +\order{\frac{1}{z}}, 
\ee
as $|z| \to \infty$. This means that the expression is bounded from below in $\C$:
\be
N\left[\frac{\alpha}{2}|z|^2+\int{\log{\frac{1}{|z-w|}}d\nu(w)}\right] \geq L 
\ee
for some constant $L \in \R$ depending on the parameters $N, \alpha$ and on the measure $\nu$. So
\be
0 \leq e^{-NV(z)} \leq e^{-\frac{N\alpha}{2}|z|^2}e^{-L},
\ee
which implies that 
\be
e^{-NV(z)} \in  L^1(\C,dm) \cap L^{\infty}(\C,dm).
\ee
Moreover,
\be
\int_{\C}|z|^k e^{-NV(z)}dm(z) \leq e^{-L}\int_{\C}|z|^k e^{-\frac{N\alpha}{2}|z|^2}dm(z) < \infty
\ee
for all $k=0,1,\dots$

\begin{flushright}
$\square$
\end{flushright}
It follows from this and the positivity of the weight that the \emph{monic orthogonal polynomials}
\be
P_{n,N}(z)  = z^n +\order{z^{n-1}} \qquad (n =0,1,\dots)
\ee
satisfying
\be
\int_{\C}{P_{k,N}(z)\c{P_{l,N}(z)}}e^{-NV(z)}dm(z) = h_{k,N}\delta_{kl}, \ k,l=0,1,\dots
\ee
exist and are unique where $h_{n,N}$ denotes the square of the $L^2$-norm of $P_{n,N}(z)$.


\section{Matrix $\c{\partial}$-problem for orthogonal polynomials}
\setcounter{equation}{0}

In this section we show that the $2 \times 2$ matrix $\c{\partial}$-problem for orthogonal polynomials introduced by Its and Takhtajan \cite{ItsTakhtajan} in the case of measures supported within a finite radius (cut-off exponentials of polynomial potentials)
is also well-defined for the class of potentials considered above and determines the polynomials uniquely. In \cite{ItsTakhtajan} the same family of potentials is considered as  in \cite{FelderElbau}. 
 
To be able to formulate the $\c{\partial}$-problem, we need some estimates of Cauchy transforms of measures with unbounded supports.
For a given potential $V(z)= V_{\alpha,\nu}(z)$ of the form (\ref{background}) considered above let $\lambda$ be the measure that is absolutely continuous with respect to the Lebesgue measure in $\C$ having the form
\be
d\lambda = e^{-NV(z)}dm
\ee
where and $N >0$. Note that $\lambda(\C)$ is finite because $e^{-NV(z)} \in L^1(\C,dm)$.

The Cauchy transform of $\lambda$ is defined to be
\be
[C\lambda](z):= \int_{\C}{\frac{d\lambda(w)}{z-w}}.
\ee

We need to control the asymptotic behaviour of the Cauchy transform $[C\lambda](z)$ of such measures at infinity allowing the possibility that $[C\lambda](z)$ is not holomorphic in any neighborhood of  $z =\infty$.

First of all, it follows from Prop. \ref{l1linfty} that the density is bounded from above: $e^{-NV(z)} < K$ for some $K \in \R$.
For a fixed positive radius $R$, we have
\bea
\ds\int_{\C}{\frac{d\lambda(w)}{|z-w|}} &\leq&\int_{|z-w|>R}{\frac{d\lambda(w)}{R}}+\int_{|z-w|\leq R}{\frac{Kdm(w)}{|z-w|}} \cr
&\leq& \frac{1}{R}\int_{\C}{d\lambda(w)}+
\int_{0}^{2\pi}\!\!\!\int_{0}^{R}{\frac{K}{r}rdrd\theta} \cr
&=& \frac{1}{R}\lambda(\C) +2\pi RK
\eea
for all $z \in \C$.
Thus, there exists an upper bound $H_{\lambda}$ of $[C\lambda](z)$ depending only on $N,\alpha, \nu$ and independent of $z$ (One can get rid of $R$ in the last expression e.g. by minimizing the bound in $R$):  
\be
\ds\int_{\C}{\frac{d\lambda(w)}{|z-w|}} < H_{\lambda} \qquad (z \in \C).
\ee

Now 
\bea
\left[C\lambda\right](z)- \frac{\lambda(\C)}{z} &=&
\int_{\C}{\left(\frac{1}{z-w}-\frac{1}{z}\right)d\lambda(w)}\cr
& = & \frac{1}{z^2}\int_{\C}{\left(\frac{wz}{z-w}\right)d\lambda(w)} \cr
& = & \frac{1}{z^2}\int_{\C}{\left(w +\frac{w^2}{z-w}\right)d\lambda(w)}.
\eea
Hence the absolute value of the difference satisfies
\bea
\left|z^2\left[\left[C\lambda\right](z)- \frac{\lambda(\C)}{z}\right]\right| &\& \leq
\int_{\C}{|w|d\lambda(w)}+ 
\int_{\C}{\frac{1}{|z-w|}|w|^2d\lambda(w)}\cr
&\& = \int_{\C}{d\hat{\lambda}(w)}+ 
\int_{\C}{\frac{d\tilde{\lambda}(w)}{|z-w|}}\leq
\hat{\lambda}(\C)+H_{\tilde{\lambda}}
\eea
where $\hat{\lambda}$ and $\tilde{\lambda}$ correspond to the measures
\be
\hat{\nu} := \nu + \frac{1}{N}\delta_0 \qquad \tilde{\nu} := \nu +\frac{2}{N}\delta_0,
\ee
respectively.
This means that
\be
\left[C\lambda\right](z)=\frac{\lambda(\C)}{z}+\order{\frac{1}{z^2}}.
\ee
If $P_{n,N}(z)$ denotes the $n$th monic orthogonal polynomial with respect to the measure $\lambda$,
the modified measure 
\be
d\lambda_n(z) := |P_{n,N}(z)|^2 d\lambda(z)
\ee
corresponds to the perturbing measure
\be
\nu_n := \nu +\frac{2}{N}\sum_{k=1}^n \delta_{a_{k}^{(n,N)}}
\ee
where $\{a_{1}^{(n,N)},a_{2}^{(n,N)},\dots, a_{n}^{(n,N)}\}$ are the zeroes of $P_{n,N}(z)$ and $h_{n,N} = \lambda_n(\C)$. An easy calculation gives
\bea
\frac{1}{z-w} &\&=\frac{1}{P_{n,N}(z)}\frac{P_{n,N}(z)-P_{n,N}(w)}{z-w}+\frac{1}{P_{n,N}(z)}\frac{P_{n,N}(w)}{z-w}\cr
&\& = \frac{1}{P_{n,N}(z)}Q_{n-1}(z,w)+\frac{1}{P_{n,N}(z)}\frac{P_{n,N}(w)}{z-w}
\eea
where $Q_n(z,w)$ is a symmetric polynomial in $z$ and $w$ of degree $n-1$ with leading order $z^{n-1}$ in the variable $z$. Therefore, by orthogonality, we get
\bea
\label{nthOP}
\int_{\C}{\frac{\c{P_{n,N}(w)}}{z-w}d\lambda(w)} &= & \frac{1}{P_{n,N}(z)}\int_{\C}{\frac{|P_{n,N}(w)|^2}{z-w}d\lambda(w)}\cr
&\& = \frac{1}{P_{n,N}(z)}\int_{\C}{\frac{d\lambda_n(w)}{z-w}}\cr
&\& = \frac{1}{z^n}\frac{z^n}{P_{n,N}(z)}\left[\frac{h_{n,N}}{z}+ \order{\frac{1}{z^2}}\right] \cr
&\& =  \frac{h_{n,N}}{z^{n+1}}+\order{\frac{1}{z^{n+2}}}.
\eea
 
Following the approach of Its and Takhtajan in \cite{ItsTakhtajan}, we consider the following $2\times 2$ matrix-valued function in the complex plane:
\be
Y_{k,N}(z) :=
\left[
\begin{array}{cc}
P_{k,N}(z) & \ds \frac{1}{\pi}\int_{\C}{\frac{\c{P_{k,N}(w)}}{w-z}e^{-NV(w)}dm(w)}\\
& \\
\ds -\frac{\pi}{h_{k-1,N}}P_{k-1,N}(z) & \ds -\frac{1}{h_{k-1,N}}\int_{\C}{\frac{\c{P_{k-1,N}(w)}}{w-z}e^{-NV(w)}dm(w)}
\end{array}
\right].
\ee
The $\c{\partial}$-derivative is
\be
\frac{\partial}{\partial \c{z}} Y_{k,N}(z) =
\left[
\begin{array}{cc}
0 & \ds -\c{P_{k,N}(z)} e^{-NV(z)}\\
& \\
0 & \ds \frac{\pi}{h_{k-1,N}}\c{P_{k-1,N}(z)} e^{-NV(z)}
\end{array}
\right]
= \c{Y_{k,N}(z)}\left[
\begin{array}{cc}
0 & - e^{-NV(z)}\\
& \\
0 & 0
\end{array}
\right].
\ee
Using the asymptotic behaviour of the Cauchy transforms as $|z|\to \infty$ proven above, we have that
\bea
&\& \left[
 \begin{array}{cc}
 P_{k,N}(z) & \ds \frac{1}{\pi}\int_{\C}{\frac{\c{P_{k,N}(w)}}{w-z}e^{-NV(w)}dm(w)}\\
& \\
\ds -\frac{\pi}{h_{k-1,N}}P_{k-1,N}(z) & \ds -\frac{1}{h_{k-1,N}}\int_{\C}{\frac{\c{P_{k-1,N}(w)}}{w-z}e^{-NV(w)}dm(w)}
\end{array}
\right]
\left[
\begin{array}{cc}
z^{-k} & 0\\
& \\
0 & z^{k}
\end{array} \right]  \cr
&\&=
\left[
\begin{array}{cc}
\ds \frac{P_{k,N}(z)}{z^k} & \ds \frac{z^k}{\pi}\int_{\C}{\frac{\c{P_{k,N}(w)}}{w-z}e^{-NV(w)}dm(w)}
\\
\ds -\frac{\pi}{h_{k-1,N}}\frac{P_{k-1,N}(z)}{z^k} & \ds -\frac{z^k}{h_{k-1,N}}\int_{\C}{\frac{\c{P_{k-1,N}(w)}}{w-z}e^{-NV(w)}dm(w)}
\end{array}
\right]  \cr
&\&=
\left[
\begin{array}{cc}
1 & 0\\
& \\
0 & 1
\end{array}
\right]
+ \order{\frac{1}{z}}.
\eea
So $Y_{k,N}(z)$ is a solution of the following $2 \times 2$ matrix $\c{\partial}$-problem:

\begin{equation}
\label{dbar}
\left\{
\begin{array}{rclc}
\ds \frac{\partial}{\partial \c{z}} M(z) & = & \c{M(z)}\left[
\begin{array}{cc}
0 & - e^{-NV(z)}\\
& \\
0 & 0
\end{array}
\right] & (z \in \C)\\
& & & \\
M(z) & = &
\left(
\left[
\begin{array}{cc}
1 & 0\\
& \\
0 & 1
\end{array}
\right]
+\ds \order{\frac{1}{z}}\right)
\left[
\begin{array}{cc}
z^k & 0\\
& \\
0 & z^{-k}
\end{array}
\right] & (|z| \to \infty).
\end{array}
\right.
\end{equation}

The important point made in \cite{ItsTakhtajan} is that the $\c{\partial}$-problem in that setting has a unique solution and therefore it characterizes the matrix $Y_{k,N}(z)$ and the corresponding orthogonal polynomials. Although we cannot assume that the relevant Cauchy transform entries are holomorphic around $z =\infty$, we nevertheless can prove that the solution is unique in this case as well. 

\begin{proposition}
The matrix $Y_{k,N}(z)$ is the unique solution of the $\c{\partial}$-problem \emph{(\ref{dbar})}.
\end{proposition}
\noindent
{\bf Proof.} We have seen that $Y_{k,N}$ solves the $\c{\partial}$-problem (\ref{dbar}). Conversely, assume that the matrix $M(z)$ has continuous entries with continuous partial derivatives and $M(z)$ solves (\ref{dbar}) with the prescribed asymptotic conditions. Then $M_{11}(z)$ and $M_{21}(z)$
are entire functions with asymptotic forms for large $z$
\bea
M_{11}(z) & = & z^k +\order{z^{k-1}},\\
M_{21}(z) & = &\order{z^{k-1}} \qquad |z| \to \infty.
\eea
Hence $M_{11}(z)$ is a monic polynomial of degree $k$ and $M_{21}(z)$ is a polynomial of degree at most $k-1$.
The $\c{\partial}$-equation in (\ref{dbar}) can be written in terms of the entries of $M(z)$ as 
\bea
\frac{\partial}{\partial \c{z}} M_{12}(z) & = & -\c{M_{11}(z)}e^{-NV(z)}, \\
\label{dbar1}
\frac{\partial}{\partial \c{z}} M_{22}(z) & = & -\c{M_{21}(z)}e^{-NV(z)}.
\label{dbar2}
\eea
Taking into account the fact that $M_{12}(z) \to 0$ and $M_{22}(z) \to 0$ as $|z|\to \infty$ this implies
\bea
M_{12}(z) & = & \ds \frac{1}{\pi}\int_{\C}\frac{\c{M_{11}(w)}}{w-z}e^{-NV(w)}dm(w), \\
\label{cau1}
M_{22}(z) & = & \ds \frac{1}{\pi}\int_{\C}\frac{\c{M_{21}(w)}}{w-z}e^{-NV(w)}dm(w)
\label{cau2}
\eea
(see \cite{Conway}).
Using the expansion
\bea
\frac{1}{z-w} &=& \frac{1}{z^k}\frac{z^k-w^k}{z-w}+\frac{1}{z^k}\frac{w^k}{z-w}\cr
&=& \sum_{l=0}^{k-1}{\frac{1}{z^{l+1}}w^l}+\frac{1}{z^k}\frac{w^k}{z-w},
\eea
we get
\bea
M_{12}(z)  &=&  \ds \frac{1}{\pi}\int_{\C}\frac{\c{M_{11}(w)}}{w-z}e^{-NV(w)}dm(w)\cr
&=& \sum_{l=0}^{k-1}{\frac{1}{z^{l+1}}\frac{1}{\pi}\int_{\C}w^l \c{M_{11}(w)}e^{-NV(w)}dm(w)}\cr
&& +\frac{1}{z^k}\frac{1}{\pi}\int_{\C}\frac{w^k\c{M_{11}(w)}}{w-z}e^{-NV(w)}dm(w).
\eea
The prescribed asymptotic behaviour 
\be
M_{12}(z)=\order{\frac{1}{z^{k+1}}} \quad \textrm{ as } |z| \to \infty
\ee
implies the following equations:
\[\int_{\C}w^l \c{M_{11}(w)}e^{-NV(w)}dm(w)=0 \qquad l=0,1,\dots,k-1.\]
Hence $M_{11}(z)=P_{k,N}(z)$ because $M_{11}(z)$ is a monic polynomial of degree $k$. Similarly for $M_{21}(z)$ we have
\bea
M_{22}(z)  &=&  \ds \frac{1}{\pi}\int_{\C}\frac{\c{M_{21}(w)}}{w-z}e^{-NV(w)}dm(w)\cr
&=& \ds \sum_{l=0}^{k-1}{\frac{1}{z^{l+1}}\frac{1}{\pi}\int_{\C}w^l \c{M_{21}(w)}e^{-NV(w)}dm(w)}\cr
& & +
\frac{1}{z^k}\frac{1}{\pi}\int_{\C}\frac{w^k\c{M_{21}(w)}}{w-z}e^{-NV(w)}dm(w)
\eea
and
\be
M_{22}(z)=\frac{1}{z^k}+\order{\frac{1}{z^{k+1}}} \quad \textrm{ as } |z| \to \infty
\ee
implies
\be
\int_{\C}w^l \c{M_{21}(w)}e^{-NV(w)}dm(w)=0 \qquad l=0,1,\dots,k-2,
\label{ortrels}
\ee
and
\be
\int_{\C}w^{k-1} \c{M_{21}(w)}e^{-NV(w)}dm(w)=1.
\ee
Now, if $M_{21}(z) = a z^{k-1} + \order{z^{k-2}}$, where $a \in \C$, then
\be
\int_{\C}\left|M_{21}(w)\right|^2 e^{-NV(w)}dm(w)=a\int_{\C}w^{k-1} \c{M_{21}(w)}e^{-NV(w)}dm(w)=a.
\ee
Clearly $a \not= 0$ (because otherwise $M_{21}(z)$ and hence also $M_{22}(z)=0$ would be zero which is impossible). So $M_{21}(z)$ is a polynomial of degree $k-1$ and from (\ref{ortrels}) we have that 
\be
M_{21}(z)=a P_{k-1,N}(z).
\ee
By the asymptotic relation (\ref{nthOP}),
\bea
M_{22}(z) & =& \ds \frac{1}{\pi}\int_{\C}\frac{\c{M_{21}(w)}}{w-z}e^{-NV(w)}dm(w)\cr
& = & \ds -\frac{a}{\pi}\int_{\C}\frac{\c{P_{k-1,N}(w)}}{z-w}e^{-NV(w)}dm(w)\cr
& = & -\frac{a h_{k-1,N}}{\pi}\frac{1}{z^{k}}+\order{\frac{1}{z^{k+1}}},
\eea
which forces the constant $a$ to be equal to $-\frac{\pi}{h_{k-1,N}}$ and hence 
\be
M_{21}(z)=-\frac{\pi}{h_{k-1,N}}P_{k-1,N}
\ee
which completes the proof.
\begin{flushright}
$\square$
\end{flushright}


\section{Zeros of Orthogonal polynomials and quadrature Domains}
\setcounter{equation}{0}

   In this final section we briefly discuss some known and conjectured relations between the asymptotics of orthogonal polynomials in the plane, equilibrium measures of the type studied in Sections 1 -- 3 and generalized quadrature domains.  These concern relations between the asymptotics of the zeros of orthogonal polynomials and the associated equilibrium measures that have previously been studied by Elbau \cite{FelderElbau, Elbau} for a certain class measures with bounded support. Their validity  for the class of measures considered here is supported by numerical calculations.
   
   To relate orthogonal polynomials with the support of the equilibrium measure we have, for finite values of $n$, three quantities which, in the case of random Hermitian matrices are known to approach the same equilibrium measure on the real line in the scaling limit (cf. (\ref{semiclassical}))
\be
\label{scaling_limit}
n \to \infty, \quad N \to \infty, \quad \frac{N}{n} \to \gamma = {1\over t},   \qquad \hbar := {1\over N} \in \R^{+}.
\ee
All the limiting relations below are understood in this scaled sense. We introduce the notation
\be
Q(z) := \frac{\gamma}{2}V(z)
\ee
for the rescaled potential corresponding to the scaling parameter $\gamma$. Then for a large class of real potentials and $z\in \Rb$, all three of the following measures converge weakly to the equilibrium measure $ d\mu_{Q} (z)$ of $Q(z)$:

\noindent
{\bf 1)} The normalized counting measure of the zeros
\be
{\mathcal Z}_{n,N} :=\{z^{(n,N)}_{1},z^{(n,N)}_{2},\dots, z^{(n,N)}_{n}\}
\ee
 of the orthogonal polynomials $P_{n,N}(z)$ with respect to the weight $\exp(-NV(z))$
\be
\nu_{n,N} := {1\over n}  \sum_{z \in {\mathcal Z}_{n,N}} \delta_{z},
\ee
\be
\nu_{n,N} \stackrel{w*}{\longrightarrow} \mu_{Q} \quad (n \to \infty).
\ee
{\bf 2)} The expected density of eigenvalues (or  \emph{one-point function}) of random Hermitian matrices
\be
\rho_{n,N}(z) ={1\over n} \sum_{k=0}^{n-1} |p_{k,N}(z)|^2 e^{-NV(z)},
\ee
drawn from the probability density
\bea
\frac{1}{Z_{n,N}} &\& \exp(-N\Tr(V(H)))dH, \\
{Z_{n,N}} &\&:= \int \exp(-N\Tr(V(H)))dH 
\eea

\be
\rho_{n,N}(z)dz \stackrel{w*}{\longrightarrow} d\mu_{Q}(z) \quad (n \to \infty).
\ee
{\bf 3)} The normalized counting measure of equilibrium point configurations
\be
{\mathcal F}^{Q}_{n} :=\{z^{Q}_1,  \dots , z^{Q}_n\}
\ee
of the two-dimensional Coulomb energy
\be
\EE_n (z_1,  \dots z_N) = {1\over 2}\sum_{i,j=1 \atop i \not=j}^N \log \frac{1}{|z_i -z_j|} + \sum_{i=1}^N Q(z_i)
\label{FiniteNenergy}
\ee
(which in the plane become the so-called \emph{weighted Fekete points})
\be
\eta_{n} =  {1\over n} \sum_{z \in {\mathcal F}^{Q}_{n}} \delta_{z}
\ee
\be
\eta_{n} \stackrel{w*}{\longrightarrow} \mu_{Q} \quad (n \to \infty).
\ee

For random normal matrices the eigenvalues are not confined to the real axis. In this case it is known \cite{HedenmalmMakarov, Elbau, AmeurHedenmalmMakarov} that in the scaling limit (\ref{scaling_limit}), the analogs of $\rho_{n,N}(z)dz$ and $\eta_{n}$ also approach the  equilibrium measure $\mu_{Q}$.
       
It is also known \cite{Elbau} that for cut-off measures of the form
\be
e^{-NV(z)}I_\DD(z), \quad V(z) = - \alpha |z|^2 + P_\harm(z), 
\ee
where $I_\DD$ is the indicator function of a compact  domain $\DD$ containing the origin whose boundary curve $\partial \DD$ is twice continuously differentiable and $P_\harm$ is a harmonic polynomial, that
\be
\lim_{N\ra \infty} \frac{1}{n}\log \frac{1}{\left|P_{n,N}(z)\right|}   =
 \int  \log \frac{1}{|z-\zeta|} d\mu_{Q}(\zeta),   \qquad z \in \C \setminus \supp(\mu_{Q}).
\ee
That is, the limit of the zeros acts effectively as an equivalent source of the external Coulomb potential.
   
  For such potentials, the boundary  $\partial(\supp(\mu_Q))$ of the support of the equilibrium measure 
  is determined through the Riemann mapping theorem as the image of the unit circle under a rational conformal map,
  whose inverse therefore has a finite number of branch points. The Schwarz function $S(z)$, defined along the boundary, determines the curve via the
  equation
  \be
\c{z} = S(z).
\ee
It  has a unique analytic continuation to the interior on the complement of  any tree $\CC^\tree$ whose vertices include the branch points,
with edges  formed form curve segments. It is  shown in \cite{Elbau} that, assuming there is a condensation limit for the orthogonal polynomial zeros supported on a tree-like graph $\CC_Q$
whose edges are curve segments, $\CC^\tree$ may be chosen so that  $\CC^Q \subset\CC^\tree$. 
Moreover, denoting by $\delta S(z)$  the jump discontinuity of $S(z)$ away from the nodes,   $\CC^\tree$  may 
be chosen as an integral curve of the direction field annihilated by the real part  $\Re[(\delta S(z)dz]$ of the  differential $(\delta S(z))dz$;
i.e. such that the tangents $X$ to the curve segments forming the edges satisfy
\be 
\Re[(\delta S(z)dz](X) =0.
\label{nullcurve}
\ee
We refer to such integral curves as {\it critical trajectories}.
 
    Based on computational evidence, and general results known for other cases 
 \cite{SavinaSterninShatalov},  there is good reason to believe that the same result holds for the class of superharmonic perturbed Gaussian measures studied in Sections 1 - 3, without the need for introducing the cutoff factor $I_\DD$. This statement, for some suitable restrictions on the permissible superharmonic perturbations, forms the first part of the conjectured relation between the zeros of the orthogonal polynomials considered in Sections 4 - 5 and the equilibrium measure $\mu_Q$.
    
 The second part gives a more detailed relation; namely, the effective density $\kappa_Q(z)$ 
 along $\CC_Q$ of the condensed orthogonal polynomial zeros is given, within a suitable scaling constant by
 \be
 d\kappa_Q(z) \sim \frac{1}{2\pi i}(\delta S(z))dz = \frac{1}{2\pi}\Im[(\delta S(z))dz].
 \ee
 Explicitly, this means that the external potential due to a uniform, normalized charge supported in $\supp(\mu_Q)$ is
 \be
 \int_{\CC_Q}\log \frac{1}{|z-\zeta|}d\kappa_Q(\zeta) = \int \log \frac{1}{|z-\zeta|} d\mu_Q(\zeta).
 \label{extpotential}
 \ee

 To support the validity of these conjectures, we take the case of the potential
\be
V(z) = \alpha|z|^2 +\beta\log{\frac{1}{|z-a|}},
\ee  
in the simply connected case considered in Section 3 above and compare the locus of the zeros of the corresponding orthogonal polynomial $P_{n,N}(z)$ with the two different integral curves of the direction field defined by (\ref{nullcurve}) joining the branch points (Fig. 6.1).
(The two other critical trajectories emanating from the branch points are omitted from the graph).

\begin{figure}[htb!]
\begin{center}
\includegraphics[width=0.7\textwidth]{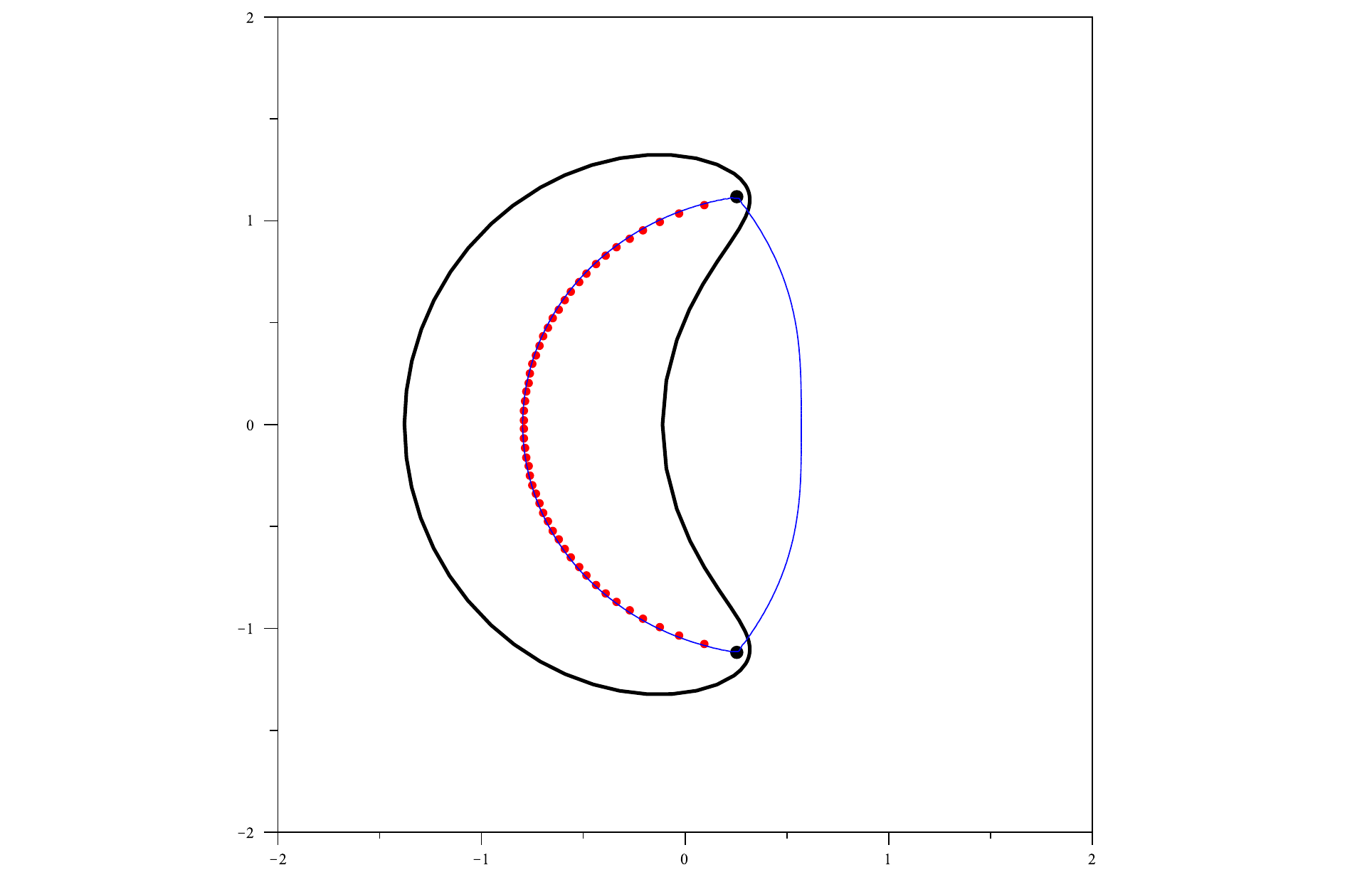}
\label{zeroes_and_traj}
\caption{Zeroes of $P_{n,N}(z)$ for $n=50$ and the critical trajectories}
\end{center}
\end{figure}

We also compare, in Fig. 6.2,  the value of the logarithmic potential
 $-{1\over n} \log|P_{n,N}(z)|$
  created by the normalized counting measure $\nu_{n,N}$ of the zeroes of $P_{n,N}(z)$ with $n=30, N=2n=60$ and the external potential as given by (\ref{extpotential}) in the external region $z \in \C\setminus \supp(\mu_V)$.

\begin{figure}[htb!]
\begin{center}
\includegraphics[width=0.4\textwidth]{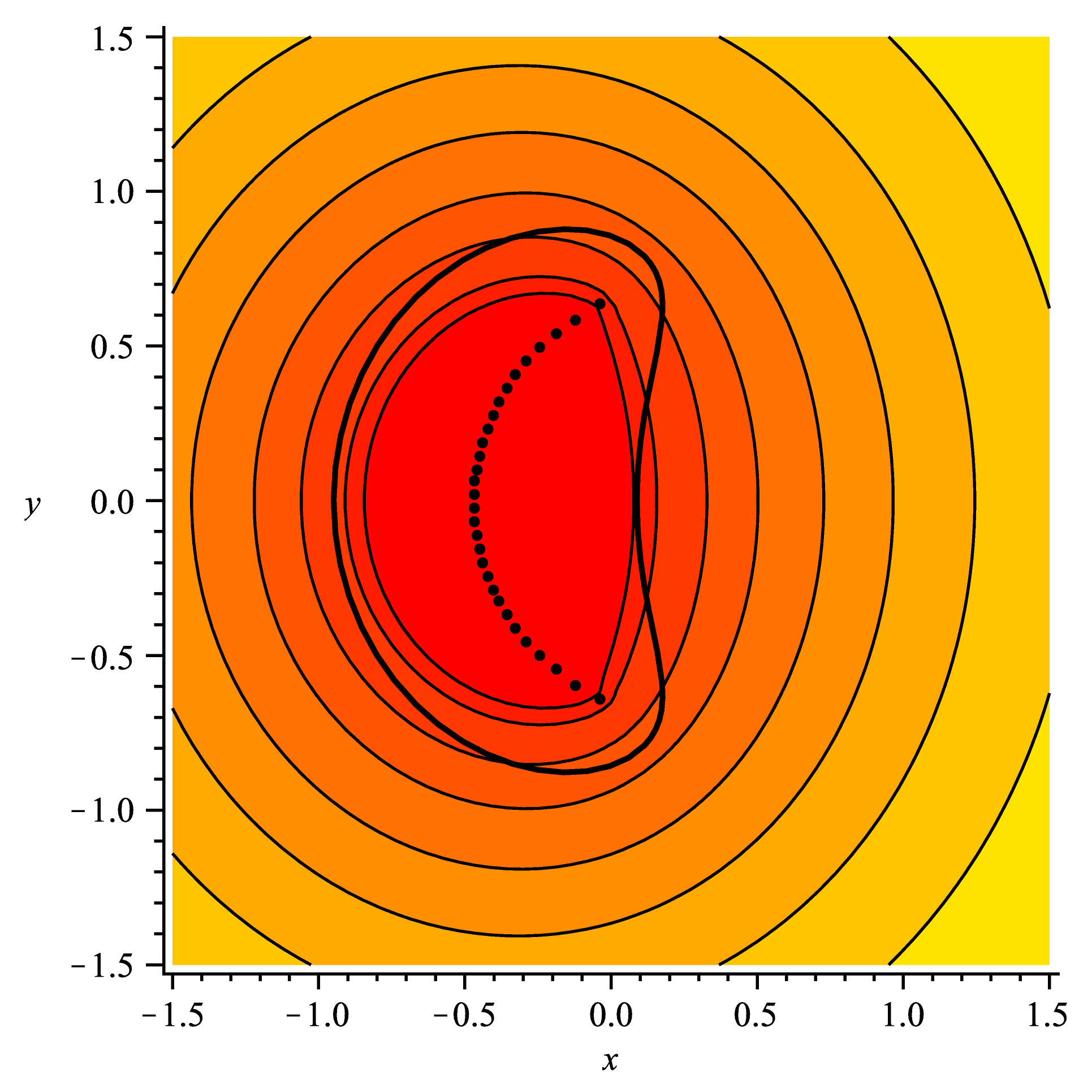}
\includegraphics[width=0.4\textwidth]{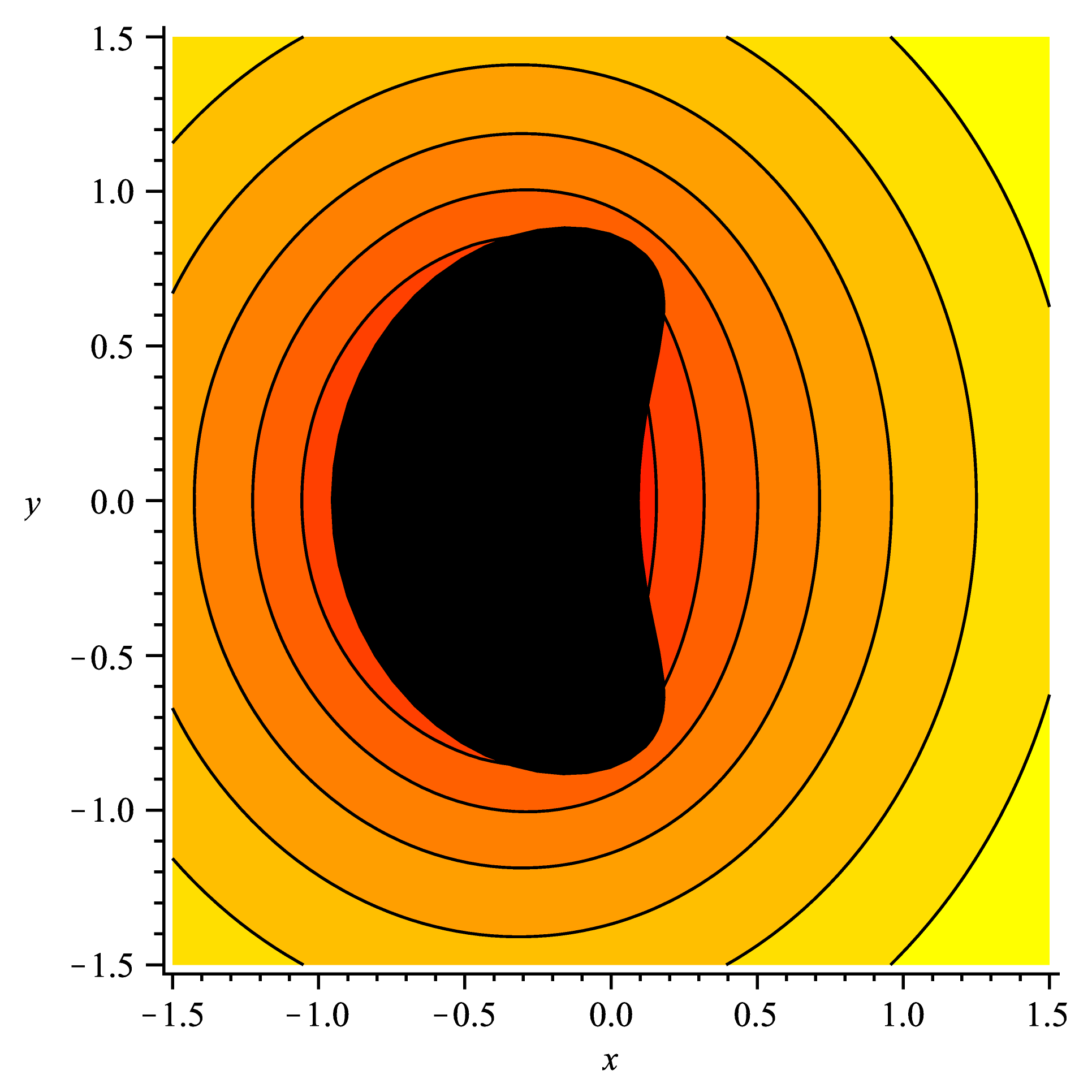}
\label{contours}
\caption{The  figure on the left is a contour plot of the potential  $- {1\over n} \log P_{n,N}(z)|$ arising from equal charges at the roots of the orthogonal polynomial
$P_{n,N}(z)$; the one on the right is the potential  $ \int \log \frac{1}{|z-\zeta|} d\mu_Q(\zeta)$ on the exterior region due to a normalized  uniform  charge on $\supp(\mu_Q)$.}
\end{center}
\end{figure}

\eject

\end{document}